\pdfoutput=1
\documentclass[aps,prl,preprint,showpacs,showkeys,groupedaddress,amsmath,amssymb]{revtex4}
\usepackage{graphicx}% Include figure files
\usepackage{dcolumn}% Align table columns on decimal point
\usepackage{bm}% bold math

\begin{document}

%\preprint{Preprint}
\title{Magnetic-field tunable photonic stop band in the three-dimensional array of conducting spheres}
\author{M. Golosovsky\footnote{e-mail golos@vms.huji.ac.il}, Y. Neve-Oz, and D. Davidov}
% \email{golos@vms.huji.ac.il}
\ \affiliation{The Racah Institute of Physics,The Hebrew University of
Jerusalem, Jerusalem 91904, Israel\\}

\date{\today} 
\begin{abstract} 
We explore possibility of tuning photonic crystal properties via order-disorder transition. We fabricated a photonic bandgap material consisting of a three-dimensional array of conducting magnetizable spheres.  The spheres self-assemble into ordered state under external magnetic field, in such a way that the  crystalline order can be continuously controlled. We study mm-wave transmission  through the array as a function of magnetic field, i.e.  for different degrees of order.  This was done for the regular crystal, as well for the crystal with the planar defect which demonstrates resonance transmission at a certain frequency. We observe that in the ordered, "crystalline" state there is a well-defined stopband, while in the completely disordered, glassy or "amorphous" state, the stopband  nearly disappears.  We relate the disappearance of the stopband in the disordered state to the fluctuations in the particle area density. We develop a model which predicts how these fluctuations depend on magnetic field and how they affect electrodynamic properties of the whole sample. The model describes our results fairly well.
\end{abstract}

\pacs {PACS numbers:  42.70.Qs, 41.20.Jb,  87.15.Ya}
\keywords{photonic  bandgap materials, tunability, self-assembly, magnetic field, wave propagation in random media, fluctuations}
\maketitle
\section{introduction}
Photonic bandgap materials are ordered arrays of scatterers that do not allow electromagnetic wave propagation in some frequency ranges  named gaps \cite{Joannopoulos}. The important challenge in the field of photonic crystals  is tunability, in other words, possibility to control the depth, width and position of these frequency  gaps.  This can be achieved by various means such as liquid crystal infiltration \cite{John},  temperature \cite{Halevi}, elastic strain \cite{Kim}, and magnetic field \cite{Lyubchanskii,Belotelov,Gates,Figotin,Xu,Xu1,Bizdoaca,Golos,Saado,Hayes}.  We explore  here a novel route to achieve tunable photonic bandgap materials: magnetic-field-induced \textit{order-disorder} transition. 
 
Since the gaps in photonic bandgap materials arise from their periodically-ordered structure, we would \textit{apriori} expect that disorder  destroys the gaps. However, it is well-known that  disorder  by itself may result in wave localization, i.e.  it may create the gap \cite{Ping}. Hence the  effect of disorder on photonic bandgap materials may be multidirectional and it comes at no surprise that it  has been studied so intensively during last years. Numerous analytical  and numerical studies showed  that the gaps in photonic crystals are robust \cite{Fan,Sigalas1,Sigalas2}, i.e. they do not disappear under weak and moderate disorder. More specifically, it was  found that  under  \textit{weak disorder}   the gap edges  become smeared \cite{Freilikher,Deych,Bayindir,Asatryan,Yannopapas,Kaliteevskii,Zhang}, the  higher-order gaps  become more shallow \cite{KP}, while the depth of the first stopband remains almost unaffected \cite{Sigalas1,Sigalas2,Deych,Bayindir,Asatryan,Kaliteevskii,KP}. Under \textit{moderate disorder} the first stopband becomes more shallow \cite{Freilikher,Zhang}. Under  \textit{strong disorder}  there appear localized states  in the gaps, which are characterized by enhanced transmission \cite{Deych,Bayindir,Kaliteevskii}.   Theoretical predictions indicate that the effect of disorder is most pronounced at the gap edges  and is minimal in the passbands,  at the frequencies corresponding to the so-called Bragg remnant or antigap \cite{KP,Sterke,Bulgakov,Kondilis}.  Various kinds of disorder (positional disorder, size disorder, fluctuations in refraction index, etc.) produce almost  the same qualitative effect. The only exception is sliding disorder (or stacking faults) which can increase the depth of the gap \cite{Asatryan,Yannopapas}.

There are only a few experimental studies of the  wave propagation in photonic bandgap materials with disorder. Ref. \cite{Bayindir} studied  microwave propagation through the  two-dimensional arrays of  rods and found  bandgap smearing  upon increasing positional disorder, while Ref.\cite{Genack} studied microwave transmission through random assemblies of spheres and observed wave localization and stopband resulting from disorder. Introduction of small controlled disorder (in fact, "quasicrystalline" order) into two-dimensional photonic bandgap material  allows to achieve a full photonic stopband in the near-infrared  \cite{Netti,Li}.  

In  this work we explore possibility of tuning of photonic bandgap materials by \textit{continuous} variation of the degree of disorder.  This is done in the mm-wave range with a model system   of  metallic magnetizable spheres whose lateral position is controlled by external magnetic field. This resembles  the polymer-dispersed  liquid crystals  where fluctuations in the optical birefringerence are controlled by electric field \cite{Levy}. A similar idea (although based on particle reorientation rather than motion in magnetic field) has been  suggested  with respect to photonic bandgap materials \cite{Figotin}. 

\section {Experimental Setup}
\subsection{Photonic crystal with tunable disorder}
We build our photonic crystal from the 2 mm diameter steel spheres. In the presence of magnetic field each sphere acquires magnetic moment proportional to the field, $M=3\chi VH/2$ where $H$ is external magnetic field,  $V$ is the  volume of the sphere, and  $\chi$ is the magnetic susceptibility. We used SQUID magnetometer to measure particle magnetization in the fields up to 500 Oe and found coercive force of  5 Oe and  volume susceptibility $\chi$=0.15.

\begin{figure}[ht]
\includegraphics[width=0.7\textwidth]{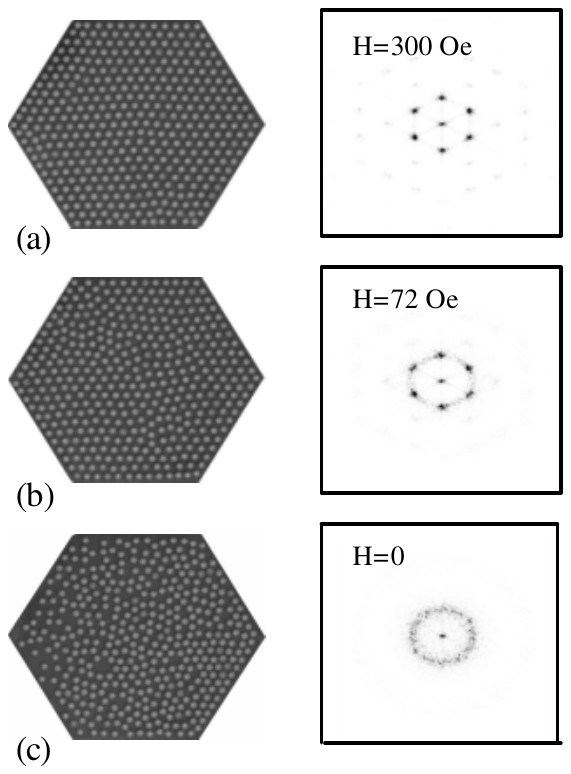}
\caption{Particle arrangement in a container at different values of magnetic field.   The container is made of a 0.67 mm thin plexiglas plate with the hexagonally-shaped walls. The  side of a hexagon is  30 mm  and there are 397 steel spheres of 2 mm diameter.  The left panel shows grey scale images of the particle configuration as obtained by the CCD camera. [The illumination  was from above, therefore the actual  diameter of the spheres exceeds the diameter of the small white circles in the real space images. In reality, the spheres in Fig. 1c touch one another.] Note gradual transition from the ordered to disordered state upon decreasing magnetic field.  The right panel shows corresponding Fourier transform image. Note gradual disappearance of sharp peaks and appearance of diffuse rings upon increasing disorder. }

\label{fig:1}
\end{figure}

\begin{figure}[ht]
\includegraphics[width=0.8\textwidth]{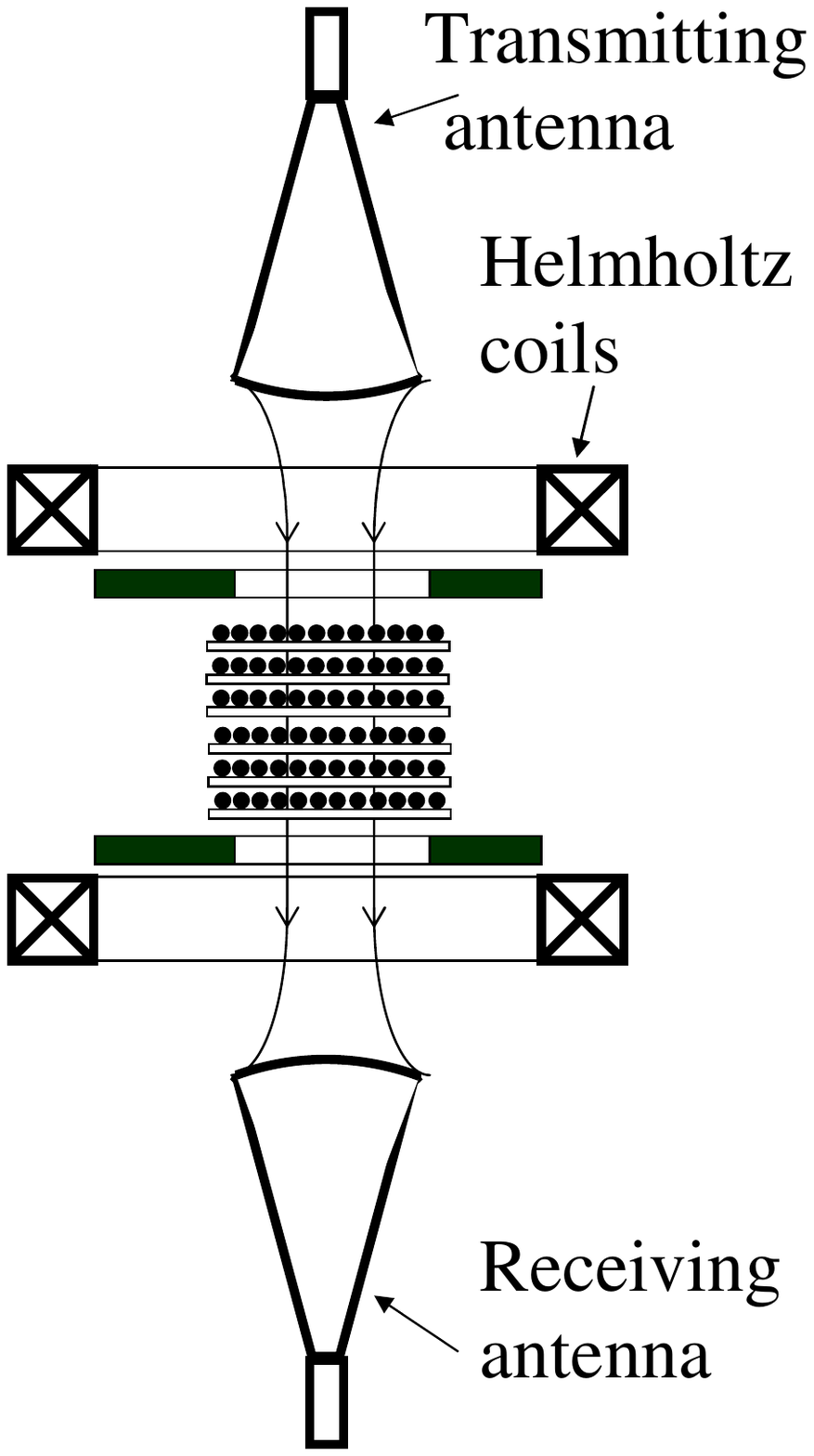}
\caption{ Measurement Setup. A stack of 6-10 layers with  steel spheres  is mounted inside Helmholtz coils which produce magnetic field  perpendicular to the layers.  The  mm-wave transmission through the stack is measured using  standard  gain microwave horns  connected to HP 8510C Vector Network Analyzer.  The antennas are equipped with collimating teflon lenses.}
\label{fig:2}
\end{figure}

We put 397 spheres into  a 0.67 mm thick  plexiglas container  (Figs.\ref{fig:1},\ref{fig:2}) and mounted several such containers  in the stack. The  interlayer   spacing  is 3.5-4.5 mm, while  the nearest-neighbor distance   in a single layer is  3 mm. The stack is mounted inside the Helmholtz coils.  In the presence of external magnetic field the  spheres  become magnetized in the  direction perpendicular to the layers,  in such a way that the spheres in each layer  repel each other.  The  out-of-plane attraction between the spheres is small  compared to the in-plane repulsion. 

The static particle configuration in each layer is determined by the interplay between (i) magnetic forces and (ii) static friction forces between the spheres and the substrate. Under \textit{strong} magnetic field, magnetic forces exceed friction forces and  the particles self-assemble into hexagonally-ordered crystalline lattice whose orientation is determined by the container shape. Since we use the hexagonally-shaped container, and  the total number of particles corresponds to the perfect hexagonal packing,   the resulting lattice is almost perfect at strong magnetic field - Fig.\ref{fig:1}. At the \textit{intermediate} magnetic field,  the friction forces deform this crystalline lattice, in such a way that it splits on several crystalline grains separated by grain boundaries.  In the \textit{weak} magnetic field, the friction forces dominate,  the particles are in the disordered (amorphous or glassy)  state  and may touch one another.
 
To characterize disorder we took images of the particle configuration using a CCD camera and then performed digital Fourier-transform of the images.  In the strong field we observe sharp Bragg peaks (Fig.\ref{fig:1}, right panel) indicating on crystalline order; at the intermediate field  the intensity of these peaks decreases  and the diffuse background appears (Fig. 1b); in the weak magnetic field the Bragg peaks disappear, indicating on the absence of the crystalline order (Fig. \ref{fig:1}). 

The images of our arrays (Fig.\ref{fig:1}) are very similar to the snapshot of the particle configuration in the system of interacting magnetic dipoles at finite temperature \cite{Ghazali}. However, instead of temperature, the disordering agent in our case is static friction between the particles and the substrate.

\subsection{Mm-wave transmission measurements}
 We measured  mm-wave transmission through our device at normal incidence and in the range of 20-50 GHz. We used a HP850C Vector Network Analyzer as a source and two standard gain horn antennae  to which we attached home-made collimating teflon lenses (Fig.\ref{fig:2}).  The inner curvature of the lens is 15 cm, the outer curvature is 30 cm, the distance between antennae is 28 cm, the beam diameter is 6 cm.  The height of the device is 4-5 cm. To prevent diffraction at the edges of the stack,  we put two apertures above and below the stack. The aperture size is slightly smaller than  the  container diameter. The receiving antenna accepts radiation in the angle of $25^{0}$, hence it measures forward transmission and  small angle scattering as well. The noise floor of our setup is -50 dB (measured by replacing a sample with a conducting foil of the same diameter).  Calibration was performed without sample. The size of the particles, the distance between them and the distance between the layers was carefully chosen in such a way as to achieve a wide and deep stopband  in the accessible frequency range. This was done by computer simulations using ANSOFT software. 

We studied transmission through the stack of 6-10 layers as a function of magnetic field, i.e. at different degrees of in-plane disorder. The measurements were performed as follows.  We applied a certain magnetic field to initially disordered array and either gently vibrated it or sent a short current pulse ("magnetic stirring") to achieve equilibration.  Then we measured mm-wave transmission through the stack. At the next step we switched off the magnetic field  and vibrated the stack once again  to recover initial, disordered state. Then we repeated the measurement for another value of magnetic field and so on. While there exist many particular  configurations corresponding to a certain value of the external magnetic field, the mm-wave transmission through these metastable configurations differ only in minor details.  Therefore, we performed 3-10 vector transmission measurements  at the same value of magnetic field  and for different particle configurations; and  vectorially averaged the results. We also checked experimentally that the transmission through the stack is almost independent of the polarization of the incident wave. A small lateral shift of one of the layers with respect to another (sliding disorder) was also found to be unimportant.  
\section{Experimental results }
\subsection{Magnetic field dependence of the mm-wave transmission} 

\begin{figure}[ht]
\includegraphics[width=0.8\textwidth]{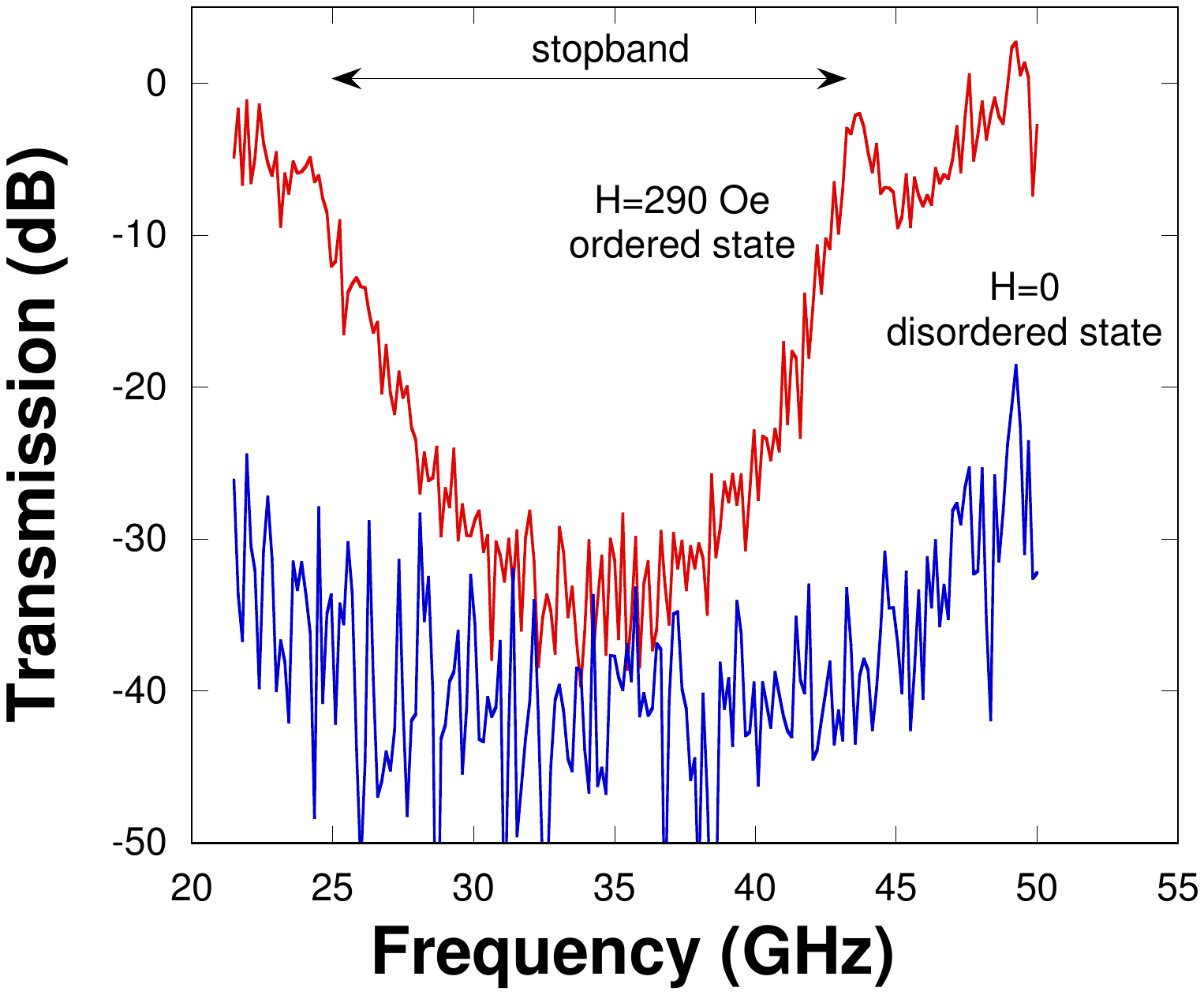}
\caption{ Mm-wave transmission through the six-layer stack  for different  values of magnetic field.   The interlayer separation  is $d=$ 3.5 mm. Note stopband at 25-44  GHz and at $H=$290 Oe  (ordered state) and its  disappearance in the disordered state ($H=$0).}
\label{fig:3}
\end{figure}

Figure \ref{fig:3} shows mm-wave transmission through the  six-layer stack  at two extreme values of  magnetic field: one corresponding to completely ordered state, and another  corresponding to completely disordered state. In the ordered state  there is a well-defined  stopband at 24-44 GHz. For the stack with only six layers, the transmission in the stopband is  very small, $T=$ 35 dB.  Reflectivity in the stopband is close to unity (not shown here) \cite{Golosovsky}. Transmission in the passband can be as high as -1-2  dB, indicating on negligible  absorption losses. This is not surprising since the losses are determined by the ratio of the skin-depth $\delta$ in the material of the sphere to its radius $r$. For the mm-wave frequencies this ratio is so small $\sim 10^{-3}$, that the spheres can be considered ideally conducting, i.e. lossless. The spikes in the stopband in the Fig.\ref{fig:3} do not arise from  the noise of the measurement system. They are related to the finite number of layers and to the resonant transmission. The spikes in the magnitude of transmission are accompanied with the similar features in the phase (not shown here).

Upon decreasing magnetic field, the stopband becomes smeared and eventually disappears.  Transmission monotonously decreases with decreasing field at all frequencies. The strongest effect of magnetic field is at the stopband edges where it achieves 30 dB; while inside the stopband the magnetic field effect is negligible. Magnetic field effect in the passband is also quite pronounced. To study it in more details we prepared a similar stack but with increased interlayer separation. Here, the  stopband is shifted to lower frequencies  in such a way  that  the frequency range above the first stopband   becomes "visible". Figure \ref{fig:4} shows the effect of magnetic field on the mm-wave transmission in this crystal. The magnetic field effect in the stopband may achieve 15 dB. Note the broad transmission peak between the first and the second stopband in the disordered state, in the frequency range  corresponding to the passband in the ordered state.   This  peak  of enhanced transmission is known as Bragg remnant or antigap \cite{Sterke,Bulgakov,Kondilis}. 

Figure \ref{fig:5} shows magnetic field dependence of the mm-wave transmission for the sample of Fig.4 and at few different frequencies, corresponding to the  midgap, gap edge, and antigap. Although the overall variation of transmission strongly depends on frequency, the field dependence is much more the same: strong linear increase below  60 Oe followed by saturation at higher field. The only exception is the midgap frequency where the overall change in transmission is so small that  its functional dependence cannot be reliably determined.

This can produce a false impression that  the mm-wave transmission is totally insensitive to magnetic field  for the frequencies corresponding to  the midgap. However, this is not exactly so. Figure \ref{fig:6} shows mm-wave transmission through the stack in which we introduced a planar defect, i.e.  we displaced two halves of the stack (Fig.\ref{fig:1}) in such a way that  the  distance between  two central layers  is increased as compared to other interlayer distances.  This results in a sharp transmission peak inside the stopband  at 34.7  GHz. Upon decreasing magnetic field this peak  decreases  and completely disappears at $H=0$. The effect of magnetic field on transmission at the resonance frequency amounts to 30 dB. Therefore, magnetic field can strongly affect transmission in the stopband provided this transmission  is resonant.

\subsection{Dispersion relations in the ordered and disordered states }
Magnetic field  affects not only the magnitude of the  transmission (Figs. \ref{fig:3},\ref{fig:4},\ref{fig:5},\ref{fig:6}) but the phase as well.  This means that the phase  and group velocity can be  magnetically  tuned. These velocities are determined from the dispersion relation $k(f)$, where $k$ is the wavevector and $f$ is the frequency.  To determine  $k(f)$ we performed vector transmission measurements at fixed field and for the stack with varying number of layers, $N$.  The real and imaginary parts of the wavevector were found from the experimental data using the following relations:
\begin{equation}
\mbox{Re}(k)=\frac{1}{d_{z}}\frac{d\phi}{d N};   \mbox{Im}(k)=\frac{1}{2d_{z}}\frac{d\ln T}{dN}\label{dispersion}
\end{equation}
where $T$ is the power transmission coefficient, $\phi$ is the phase shift upon transmission, $d_z$ is the unit cell period in the direction of propagation. Figure \ref{fig:7} shows our results. In the ordered state, the $\mbox{Im}(k)$  is very small beyond the stopband, as expected in the lossless material. This is in contrast to the disordered state where the $\mbox{Im}(k)$ is more or less the same at all frequencies. Note the gap in $\mbox{Re}(k)$ in the ordered state and the absence of the gap in the disordered state.  The group velocity is $v_{g}=dk/df$. In the disordered state it is fairly frequency-independent, while in  the ordered state it demonstrates strong frequency dependence, in particular, $v_{g}\rightarrow 0$ at the stopband edges. Therefore, group velocity can be  continuously tuned  by magnetic field.

\section{Modeling}
Although particle configuration in each layer strongly depends on magnetic field,  to understand how  the changes in configuration  affect  the mm-wave transmission through the whole sample is not an easy task. Indeed,  to explain the appearance of the stopband  for the certain direction of propagation in a three-dimensional photonic bandgap material, it is usually enough to  represent it as a  multilayer with  periodicity \textit{only}  in the direction of propagation. In this representation the layers are assumed spatially uniform and the details of  the  particle arrangement in each layer are of little importance. Our idea is that any deviation from the in-plane crystalline order  leads to fluctuations in the area density of particles. Therefore, the layer can be no more considered as uniform.  

The purpose of our modeling is to find how  the  \textit{in-layer} disorder affects the electromagnetic wave propagation \textit{perpendicular} to the layers. We attribute the effect of lateral disorder to the area density fluctuations.  The modeling is performed  in three steps. First we estimate fluctuations of the particle area density  as a function of magnetic field. Secondly, we calculate refraction index and reflectivity of a single layer of spheres with fluctuating area density.  Thirdly, we consider wave propagation through the stack of such layers.

\subsection{Magnetic-field-dependent disorder in a  layer of magnetizable spheres}
Particle configuration  in our array is determined by the interplay between magnetic and friction forces.  Consider first magnetic forces.  The potential energy of induced  magnetic dipole-dipole repulsion is:
\begin{equation}
U_{ij}=\frac{M_{i}M_{j}}{|r_{i}-r_{j}|^{3}}\label{Hamiltonian}
\end{equation}%Eq.2
where $r_{i}$ is the particle position,  $M_{i}$ is the particle magnetic moment.  Magnetic forces  $F^{i}_{magn}=\sum_{j}\frac{\partial U_{ij}}{\partial r_{i}}$  drive particles into equilibrium hexagonally-ordered state  where magnetic repulsion energy is minimized and  $F^{i}_{magn}=0$. The lattice constant of this array is determined by the number of particles and the size and shape of container \cite{Saado}. Since magnetic susceptibility of our spheres is not high,  we assume $M\propto H$ where $H$ is \textit{external} field (rather than the sum of the external field  and the induced field of other particles).

To achieve equilibrium, the particles should move laterally. Here,  static friction forces become important. Indeed, to push a hard sphere  on flat substrate out of equilibrium requires application of some mimimal force, exceeding static friction force, $F_{fr}=\mu_{fr}P$ where $P$ is normal force on a particle and $\mu_{fr}$ is the rolling friction coefficient which depends on the radius of the sphere and on the nature of contacting surfaces. In the presence of static  friction  the particles start to move only when  $F_{magn}> F_{fr}$ and come to rest  when $F_{magn}\leq F_{fr}$.     
To characterize the interplay between magnetic and friction forces  we introduce "magnetic length" $2l_{H}$  that represents a minimal distance between two isolated magnetized spheres  on  substrate with friction. We find it  from the relation $F_{magn}(2l_{H})=F_{fr}$. Equation (\ref{Hamiltonian}) yields \begin{equation}
l_{H}=\left(\frac{3M^{2}}{16F_{fr}}\right)^{1/4}\label{l-H}
\end{equation}
In this work we are interested in the magnetizable particles ($M\propto H$), hence $l_{H}\propto H^{1/2}$. The magnetic length should be compared to the  average distance between the particles. The latter is conveniently characterized by  the Wigner-Seitz radius $a=(\pi\rho)^{-1/2}$, where $\rho$ is the area density of the particles.  Note that $l_{H}$ is controlled by magnetic field (Eq.\ref{l-H}) while $a$ is field-independent. In what follows we consider different regimes, defined by the relation between $l_{H}$ and $a$.

\subsubsection{Strong magnetic field (crystalline state)} 
Here $l_{H}>>a$. Magnetic repulsion  tends to keep the particles at equal distances. For confined array and at high enough field this results in the hexagonally-ordered "crystalline" state.   The whole array is a single crystallite (with probably few point defects) which is deformed by friction forces. In the absence of point defects such as dislocations and disclinations \cite{Kusner,Golos-synt},  we estimate the area density fluctuations  arising from elastic deformations as follows.  The condition of static equilibrium of the elastically  deformed planar crystalline lattice is
\begin{equation}
\frac{\partial \sigma_{ik}}{\partial x_{k}}+\rho f_{i}=0\label{equil}
\end{equation}
Here $\sigma_{ik}$ is the stress tensor and $f_{i}$ is the density of bulk external forces which in our case  are friction forces. The  magnitude of these forces  is  $F_{fr}$.  We assume that their correlation radius  is the Wigner-Seitz radius, $a$. By integrating Eq.\ref{equil} over unit cell, we find $|\sigma_{ik}|\approx \rho F_{fr}a$. The compressional deformations are $|u_{ll}|\sim \rho F_{fr}a/K$, while the shear deformations are $|u_{ik}|\sim \rho F_{fr}a/\mu, (i\neq k)$. Here, $K$ is the bulk compression modulus and $\mu$ is the shear modulus. For the planar hexagonal lattice of parallel magnetic dipoles  \cite{Kusner}
\begin{equation}
 \mu=\frac{2.147M^{2}}{a^5}; K=10\mu\label{mu}
 \end{equation}

Equations \ref{l-H} and \ref{mu} yield  fluctuations of the deformations, $|u_{ll}|\approx 2.75\times 10^{-3}(a/l_{H})^{4}$; $|u_{ik}|\approx 10|u_{ll}|$.  The corresponding density fluctuations  are very small:
\begin{equation}
\frac{\overline{\delta \rho^2}}{\rho^{2}}\sim \overline{u_{ll}^2}\propto\left(\frac{a}{l_{H}}\right)^{8}\propto H^{-4}\label {fluct-uni}
\end{equation}

\subsubsection{Intermediate field (polycrystalline state)}
Here $l_{H}>a$.  In this regime, the friction forces between the particles and the substrate  are strong enough to  split the crystalline lattice onto separate grains with well-defined grain boundaries.  To estimate the grain size $R$, we follow  Larkin-Ovchinnikov treatment of the pinned vortex lattice in superconducting films \cite{LO}.  The maximum energy, associated with one particle in the  lattice pinned by friction force, is $\sim F_{fr}a$. Since the friction forces are randomly oriented, the average pinning energy for the grain of area $\pi R^2$, containing $Z_{g}=(R/a)^2$ particles, is $\sim F_{fr}aZ_{g}^{1/2}= F_{fr}R$. These pinning forces lead to displacement of the grain boundary by $\sim a$. The elastic energy associated with this displacement is $\sim\mu a^2$.  The excess  energy per unit area of such lattice is:
 \begin{equation}
\frac{\delta E}{S}\sim\frac{\mu a^{2}-F_{fr}R}{\pi R^2}\label{LO}
\end{equation}
Since $K>>\mu$ we neglect here  the elastic energy due to compressional deformations. Minimization of Eq.\ref{LO} yields the equilibrium grain size, $R=2\mu a^{2}/F_{fr}$.  Using  Eqs.\ref{l-H},\ref{mu} we find the grain size  
\begin{equation}
R\approx 22.4 \frac{l_{H}^{4}}{a^3}\propto H^2\label{R}
\end{equation}
%Note that the crystalline grain  should comprise many particles. Since at the intermediate field, $l_{H}>a$, then  Eq. \ref{R} indeed yields $R/a>1.4$. This proves self-consistency of our approach. 

To estimate density fluctuations we note that the lattice in the grains is almost perfect, while the particle configuration  at grain boundaries is  distorted and strong deformations $u_{ik}$ occur there (Fig.\ref{fig:1}). In other words,  $u_{ik}<<1$ in the grains and $u_{ik}\sim 1$ at grain boundaries. The relative concentration of grain boundary particles is $Z_{B}\approx 2a/R$.  The  average fluctuation  of the deformations in the area $S$,  containing $Z_{S}=\rho S$ particles, is 
\begin{equation}
 \overline{u_{ik}^2}\sim\frac{Z_{B}}{Z_{S}}\sim \frac{0.09}{Z_{S}}\left(\frac{a}{l_{H}}\right)^4\propto {H^{-2}}\label {fluctLO}
\end{equation}
%Since $l_{H}\propto H^{1/2}$, then  $\overline{\delta \rho^2}\propto 1/H^2$.
 
%Self-consistency requires that the area density fluctuations in the crystalline state should be small compared to those in amorphous state.  Indeed, at $l_{H}<1.2a$ (disordered state) the Eq.\ref{fluct-low} yields larger fluctuations than  Eq. \ref{fluctLO}   yields at $l_{H}>1.2a$ (ordered state).

\subsubsection{Weak magnetic field (amorphous state)}
Here  $l_{H}<< a$. The  particle arrangement  reminds a glass or amorphous solid consisting of impenetrable spheres with the radius $l_{H}$.  The  area density fluctuation in the area $S$ containing $Z_{S}$ particles is \cite{LL}:
\begin{equation}
\frac{\overline{\delta \rho^2}}{\rho^{2}}=\frac{\overline{\Delta Z_{S}^2}}{Z_{S}^2} =\frac{1+\int \nu \partial S}{Z_{S}}\label {fluct2}
\end{equation}
where   $\nu$ is the pair correlation function. Following Ref.\cite{Portis}, we assume that the correlation function for the  dense  disordered planar array of impenetrable spheres of radius $r$ is such that $\int \nu \partial S\approx-\pi \rho r^2$. We substitute $r$ by $l_{H}$, $\rho$ by $1/\pi a^{2}$, introduce these values into Eq.\ref{fluct2} and find  
\begin{equation}
\frac{\overline{\delta \rho^2}}{\rho^{2}}=\frac{1-\frac{ l_{H}^2}{a^{2}}}{Z_{S}}\label {fluct-low}
\end{equation}
 Since $l_{H}\propto H^{1/2}$,  the Eq.\ref{fluct-low} yields   density fluctuations \textit{linearly}  decreasing with $H$. Note, that if  $l_{H}<r$, where $r$ is the radius of the  sphere, the magnetic forces are too small as compared to friction forces, and magnetic field is  inoperative.

\subsection{Electrodynamic parameters of a  planar array of  conducting spheres}
We consider first an \textit{ordered} planar array of  ideally conducting spheres of radius $r$, such that  $r<\lambda/2\pi$, where  $\lambda$ is the wavelength of the incident wave (such spheres can be considered as  Rayleigh scatterers). The average nearest-neighbor distance is smaller than $\lambda/2$.  When the planar electromagnetic wave is incident on such a layer, it is not absorbed but scattered.  The scattering occurs mostly in the   backward and forward  directions, while the scattering  in oblique directions is strongly suppressed. Therefore, for the  normally incident plane wave this layer can be considered as a \textit{uniform} non-absorbing medium with thickness $d=2r$, effective  refraction index $n$ and admittance $Y$ \cite{Tretyakov}. For dilute arrays of scatterers the Refs. \cite{Hulst,Bohren} yield
\begin{equation}
n\simeq 1-i2\pi\frac{\rho S(0)}{k^{3}d}\label{n}
\end{equation}
\begin{equation}
Y\simeq 1-i2\pi\frac{\rho S(180)}{k^{3}d}\label{Y}
\end{equation}
Here, $\rho$ is the area density of particles; and  admittance $Y$  is normalized to the admittance of free space. The  $S(0)$ and $S(180)$ are the forward and backward scattering amplitude of a single  particle, correspondingly, which are related to their electric $\alpha_{E}$ and magnetic $\alpha_{H}$ susceptibilities  as follows \cite{Hulst,Bohren}:
\begin{equation}
S(0)=ik^{3}(\alpha_{E}+\alpha_{H}), S(180)=ik^{3}(\alpha_{E}-\alpha_{H})\label{S}
\end{equation}
The susceptibilities depend on particle concentration due to Lorentz field. Indeed,  the susceptibility of a polarizable dipole in  the array of identical dipoles is 
\begin{equation}
\alpha=\frac{\alpha_{0}}{1-A\frac{\alpha_{0}}{ a^{3}}}\label{Lorentz}
\end{equation}
Here $\alpha_{0}$ is the susceptibility of an isolated particle (for a small isolated ideally conducting sphere,  $\alpha_{E}=r^{3},\alpha_{H}=-r^{3}/2$); $a=(\pi\rho)^{-1/2}$ is the Wigner-Seitz radius, and  $A$ is a  local field factor \cite{A}.  To find refraction index and admittance of a planar \textit{ordered} array of ideally conducting spheres,  we substitute Eq.\ref{Lorentz} into Eqs. \ref{n},\ref{Y},\ref{S} and find 
\begin{equation}
n\simeq1+ \frac{ x^{2}(1+2Ax^3)}{(1-Ax^3)(2+Ax^{3})}\label{n1}
\end{equation}
\begin{equation}
Y\simeq1+ \frac{3 x^{2}}{(1-Ax^3)(2+Ax^{3})}\label{Y1}
\end{equation}
where $x=r/a$.   Note that  $n,Y$ depend on density (through $x$ and $a$). 

To find refraction index and admittance of the  planar \textit{disordered} array of ideally conducting spheres  we represent it by the ordered array with the spatially-varying density $\rho=\rho_{0}+\delta\rho$. The Taylor expansion of Eqs.\ref{n1},\ref{Y1} with respect to $\rho$ yields 
\begin{equation}
n(\rho)\simeq n(\rho_{0})+ \frac{d n}{d\rho}\delta \rho +\frac{d^{2} n}{d\rho^{2}}\frac{\delta \rho ^{2}}{2}...\label{n2}
\end{equation}
\begin{equation}
Y(\rho)\simeq Y(\rho_{0})+ \frac{dY}{d\rho}\delta \rho +\frac{d^{2} Y}{d\rho^{2}}\frac{\delta \rho ^{2}}{2}...\label{Y2}
\end{equation}
To find $n,Y$ in the disordered state we use  the same Eqs. \ref{n2},\ref{Y2} and substitute  $\delta \rho ^{2}$ by area density fluctuations. Then we average over the area $S$ which scatters coherently.  Its size is $S=z\lambda$ where $z$ is the distance to the observation point and $\lambda$ is the wavelength. (When the averaging is performed over whole array,   $<\delta\rho>=0$, but $<\delta\rho^2>\neq 0$). 

The above approach reduces the effect of  disorder on $n,Y$ to the area density fluctuations. When the array of particles is ordered, this corresponds to  compressional deformations. However, disorder introduces  shear deformations as well. These deformations change local crystalline symmetry and this affects refraction index and admittance through the  local field factor $A$ \cite{Portis,A}. In principle, the effect of shear deformations may be treated along the same lines by considering Taylor expansion of Eqs.\ref{n1},\ref{Y1} with respect to $A$.

\subsection{Electrodynamic parameters of the  multilayer with in-plane disorder}
We represent our sample as a multilayer consisting of alternating layers of conducting spheres and air spacings between them.  Wave transmission through  multilayers is most easily accounted  for by the matrix method \cite{Pendry,Yeh}. Here, each layer is characterized by two matrices: (i) the phase matrix representing the phase shift upon transmission through this layer, and  (ii) the reflectivity  matrix which is determined by the ratio of admittances  at the reflecting interface and is  independent of the layer thickness. Transmission through the multilayer is the product of all these matrices. We relate the effect of disorder on transmission through our multilayer to the area density fluctuations in the layers of conducting spheres. We assume that the lateral size of important fluctuations  corresponds to the area that radiates  coherently, \textit{as seen from the next interface}, i.e. $S\sim\lambda d_{z}$ where $d_{z}$ is the unit cell period in the direction of propagation and $\lambda$ is the wavelength.  In what follows we consider the effect of area density fluctuations  in different frequency ranges. 

\subsubsection{Antigap} The antigap is the narrow frequency range in the passband  where the layer  of spheres may be represented as a half-wavelength plate, i.e., the phase shift on propagation through this layer is a muliple of $\pi$.   Transmission  through  such layer is close to unity  due to destructive interference of the waves reflected from  both interfaces and is almost independent on the layer admittance. Transmission at the antigap frequency is, therefore, determined mostly by refraction index fluctuations across the layers (Eq.\ref{n2}). The latter  result  in the  phase shift fluctuations, $\psi=\psi_{0}+\delta\psi$.  The power transmission through the multilayer decreases by the Debye-Waller factor, $T_{disorder}/T_{order}=e^{-\overline{\delta\psi^2}}$ \cite{LL1}. What is important here is the relative deviation of the phase shift when going from one layer to another. Hence, the averaging in the exponent is for different layers, i.e. for different realizations of the disordered state. Since $\psi=k_{0}nd$, where  $k_{0}$ is the free space wavevector, then $\delta\psi= k_{0}d\delta n=k_{0}d\frac{d n}{d\rho}\delta\rho$. Hence,  the ratio of transmittances for the $N$-layer stack  is  
\begin{equation}
\left ( \ln \frac{T_{disorder}}{T_{order}}\right)_{antigap}=-N\left(k_{0}d\frac{d n}{d\rho}\right)^{2}\overline{\delta\rho^2}\label{antigap}
\end{equation}
Note that the ratio of transmittances  is $\propto k_{0}^2$, i.e. increases with frequency.  Therefore the effect of disorder is more pronounced  in the higher-order passbands.

\subsubsection{Midgap} 
In the midgap the reflected waves from all interfaces sum up in phase, hence the reflectivity is maximal and transmission is minimal. For simplicity, we restrict ourselves to the first stopband and to the quarter-wavelength stack.  Transmission through such  stack  is   $T\approx 1/Y^{2N}$ \cite{Yeh} where $N$ is the number of layers. The average admittance of each layer in the disordered state is $Y_{disorder}=Y_{order}+\frac{d^2 Y}{d\rho^2} \overline{\delta\rho^2}$ where the averaging is performed over the whole layer.   This results in 
\begin{equation}
\left ( \ln \frac{T_{disorder}}{T_{order}}\right)_{midgap}=-\frac{N}{Y}\frac{d^2 Y}{d\rho^2} \overline{\delta\rho^2}\label{midgap}
\end{equation}

\subsubsection{Stopband edges}  
To account for the effect of disorder on the transmission at gap edges we draw  analogy to the optical properties of disordered semiconductors. Upon increasing disorder, the gap in semiconductors  becomes smeared  and there appear exponential transmission tails at gap edges (Urbach tails). The frequency dependence of optical transmission through semiconductors and at the gap edge is \cite{Goldstein}:
\begin{equation}
\frac{\partial \ln T}{\partial f}\propto\overline{\delta a_{L}^{2}}\label{urbach}
\end{equation}
where $\overline{\delta a_{L}^{2}}$ is the lattice constant variation which  can arise  from structural or temperature fluctuations. We rewrite Eq.\ref{urbach} for a single frequency, and note that $\delta a_{L}/a_{L}=\delta\rho/2\rho $. Then Eq.\ref{urbach} yields the ratio of the transmissions in the ordered and disordered states at any  frequency corresponding to the gap edge as
\begin{equation}
\left ( \ln \frac{T_{disorder}}{T_{order}}\right)_{edge}\propto\overline{\delta\rho^{2}}\label{urbach1}
\end{equation}

\subsubsection{Scattering}
In addition to the frequency-selective effects considered in previous subsections, there is also scattering that decreases transmission at all frequencies \cite{Kaliteevskii,Romanov}. These extinction losses are also proportional to density fluctuations \cite {Hulst}:
\begin{equation}
\left(\ln\frac{T_{disorder}}{T_{order}}\right)_{scattering}\propto\overline{\delta\rho^{2}}\label{scattering}
\end{equation}

We observe that   Eqs.\ref{antigap},\ref{midgap},\ref{urbach},\ref{scattering} - all yield exponential dependence of the transmission on density fluctuations, 
\begin{equation}
\ln \frac{T(f,H)}{T_{ord}}= -B(f)\frac{\overline{\delta\rho^{2} (f,H)}}{\rho^{2}}\label{master}
\end{equation}
where $T_{ord}$ is the transmission in the completely ordered state and $B(f)$ is the frequency-dependent prefactor.   Equation \ref{master} assumes monotonous dependence of transmission on disorder. This is valid only for the relatively  weak disorder, while for the strong disorder this dependence can  be nonmonotonous in certain frequency ranges \cite{Freilikher}.

\section{Comparison to experiment}
\subsection{Magnetic-field dependent order-disorder transition}
We explore magnetic field dependence of the in-plane disorder. We quantify  disorder through  the intensity of  diffuse rings in the Fourier transform images  (Fig.\ref{fig:1}).  As it is well-known from the $X$-ray structure analysis, $I_{diff}$ is directly related to the lattice constant fluctuations,  which are closely related to the density fluctuations,  i.e. $I_{diff}\propto \overline{ \delta \rho^{2}}/\rho^2$. At Figs.\ref{fig:8},\ref{fig:9} we plot $I_{diff}$ in dependence of magnetic field. Here, we set $I_{diff}=1$  in the completely disordered state, i.e. at $H=0$. In the weak field  the data  can be well approximated by the linear dependence,  $I_{diff}=1-\beta H$  as predicted by Eq.\ref{fluct-low} for the amorphous state. At higher field, the data are fairly well approximated by the $1/H^2$ dependence (see also Fig.\ref{fig:9}), as predicted by Eq.\ref{fluctLO} for the polycrystalline state \cite{Synt}.  The crossover between these two regimes occurs at $H=$ 50-60 Oe. This  corresponds to the field when the Bragg peaks in the Fourier transform image are replaced by diffuse rings (Fig.1) and is suggestive of some kind of melting or glass transition \cite{Zahn,Bubeck,Lozovik,Weiss,Peeters}. 

\subsection{Mm-wave transmission through the regular crystal}
To verify relation between transmission  and area density fluctuations, as predicted by Eq.\ref{master}, we plot them  together  versus  magnetic field (Figs.\ref{fig:8},\ref{fig:9}).  Both $I_{diff} (H)$ and $T(H)$ dependences collapse on one curve. This means that $T(H)$ is indeed proportional to $I_{diff}$ and to the density fluctuations, as predicted by Eq.\ref{master}. \textit{This is the central result of our paper}.

The absolute value of the fluctuations in the completely disordered state may be estimated as follows. The Eq. \ref{fluct-low} predicts that at $H=0$, $\overline{\delta\rho^{2}}/\rho^2\approx 1/Z_{S}$. Since $Z_{S}= \rho d_{z}\lambda$, we substitute $\rho=0.17 mm^{-2}$, $d_{z}=4.5 mm$, $\lambda=$ 6-15 mm, and find $Z_{S}\approx 5-11$ depending on frequency. This corresponds to $\overline{\delta\rho^{2}}/\rho^2=0.1-0.2$. 

\begin{figure}[ht]
\includegraphics[width=0.8\textwidth]{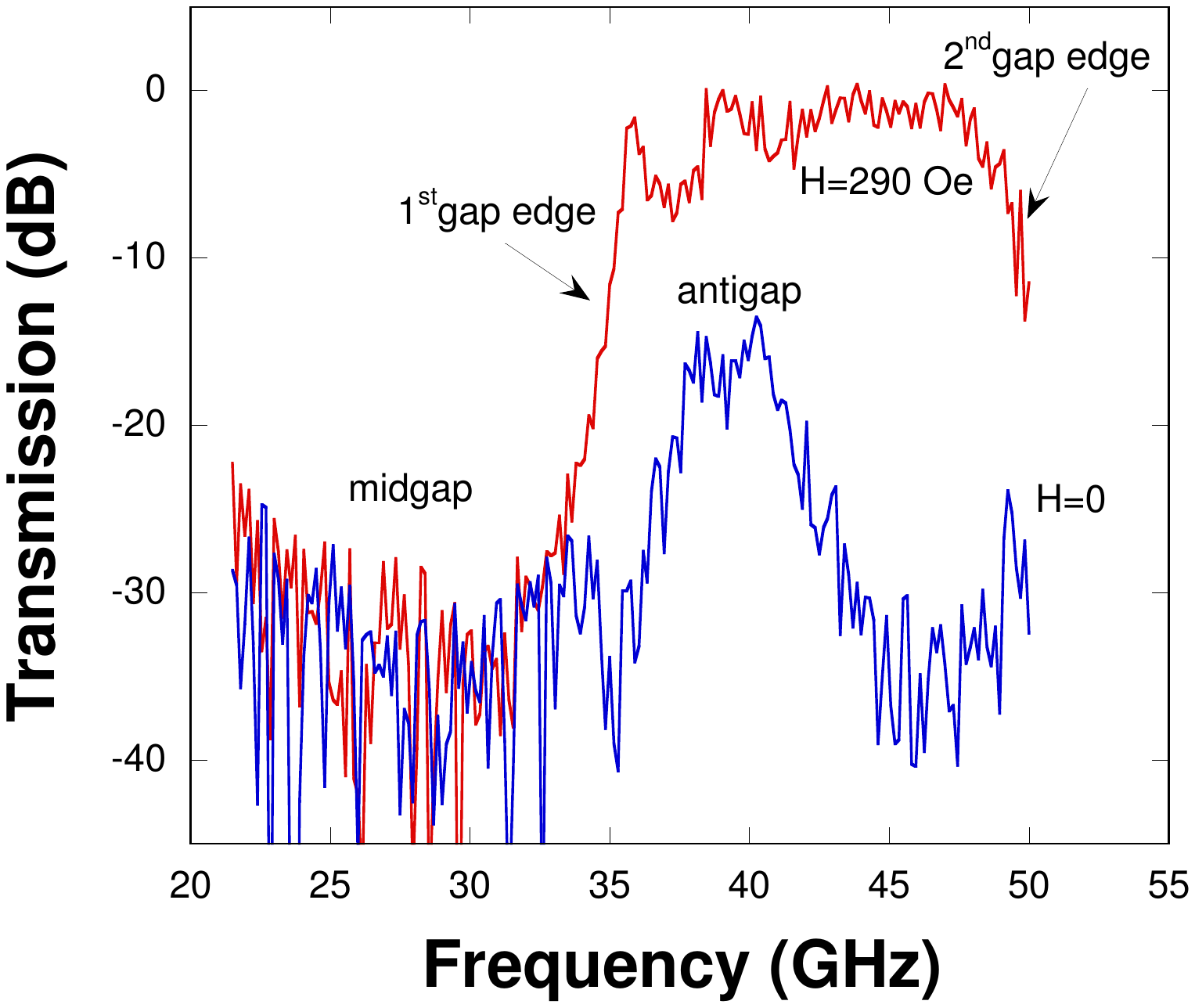}
\caption{ Mm-wave transmission  for the similar stack but with increased interlayer distance  ($d=$ 4.5 mm). In the ordered state ($H=$290 Oe) note first stopband at 21-36 GHz and the edge of the second stopband at 46 GHz. Note smearing of the stopband in the disordered state and broad transmission peak at  39 GHz  (antigap). }
\label{fig:4}
\end{figure}

\begin{figure}[ht]
\includegraphics[width=0.8\textwidth]{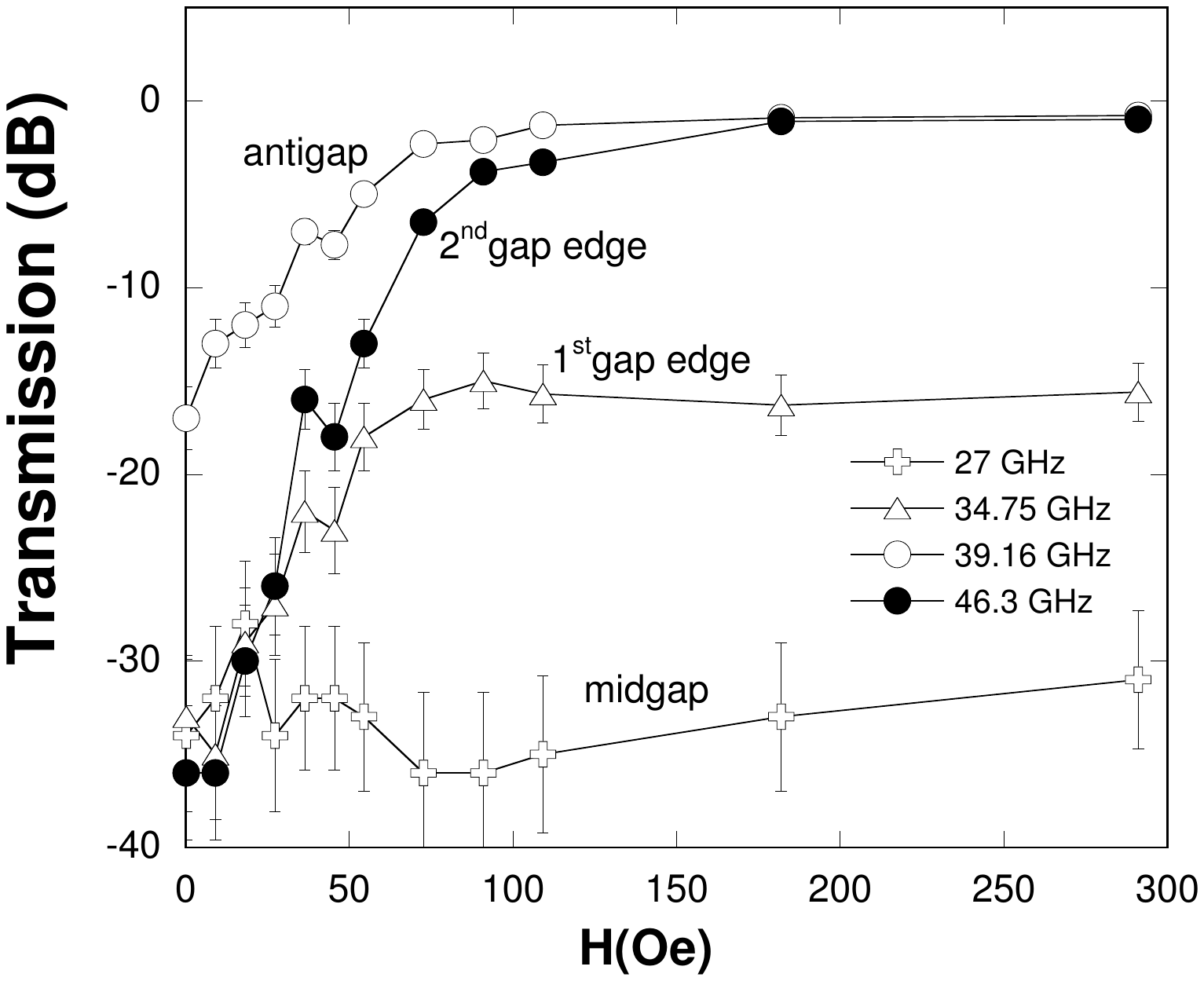}
\caption{Magnetic field dependence of the mm-wave transmission  through the sample of Fig.4 and at fixed frequencies.  $f=$34.7GHz corresponds to the high-frequency edge of the first stopband, $f=$39.1 GHz corresponds to the antigap, and $f=$46.3 GHz corresponds to the low-frequency edge of the second stopband. }
\label{fig:5}
\end{figure}

\begin{figure}[ht]
\includegraphics[width=0.8\textwidth]{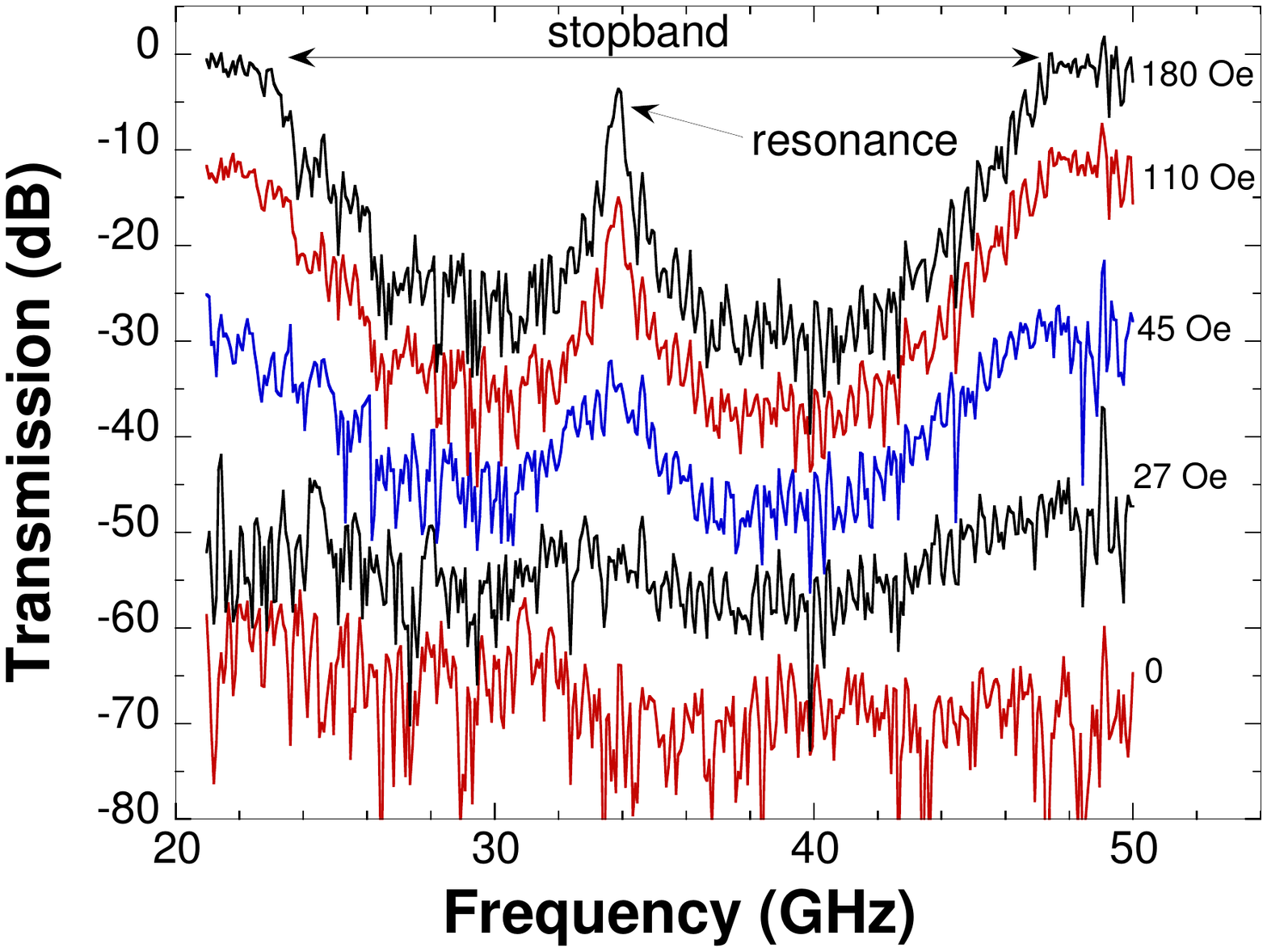}
\caption{Mm-wave transmission  for the same stack as in Fig.\ref{fig:3} but  with planar defect (the stack shown in Fig.\ref{fig:1} was split in two parts  separated by 6 mm).  Note  a sharp peak at 34 GHz inside the stopband which corresponds to the resonant transmission.}
\label{fig:6}
\end{figure}

\begin{figure}[ht]
\includegraphics[width=0.4\textwidth]{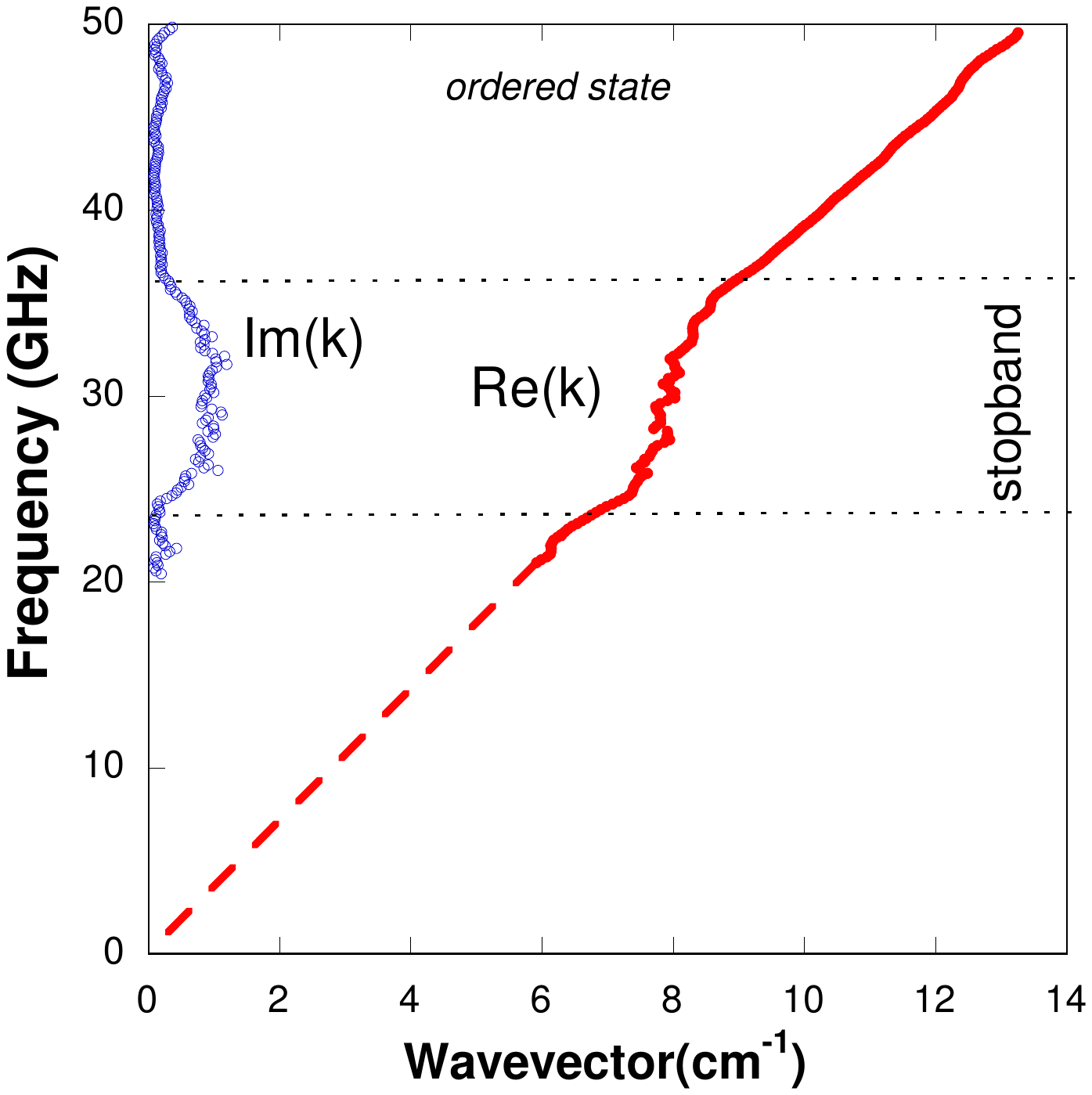}
\includegraphics[width=0.4\textwidth]{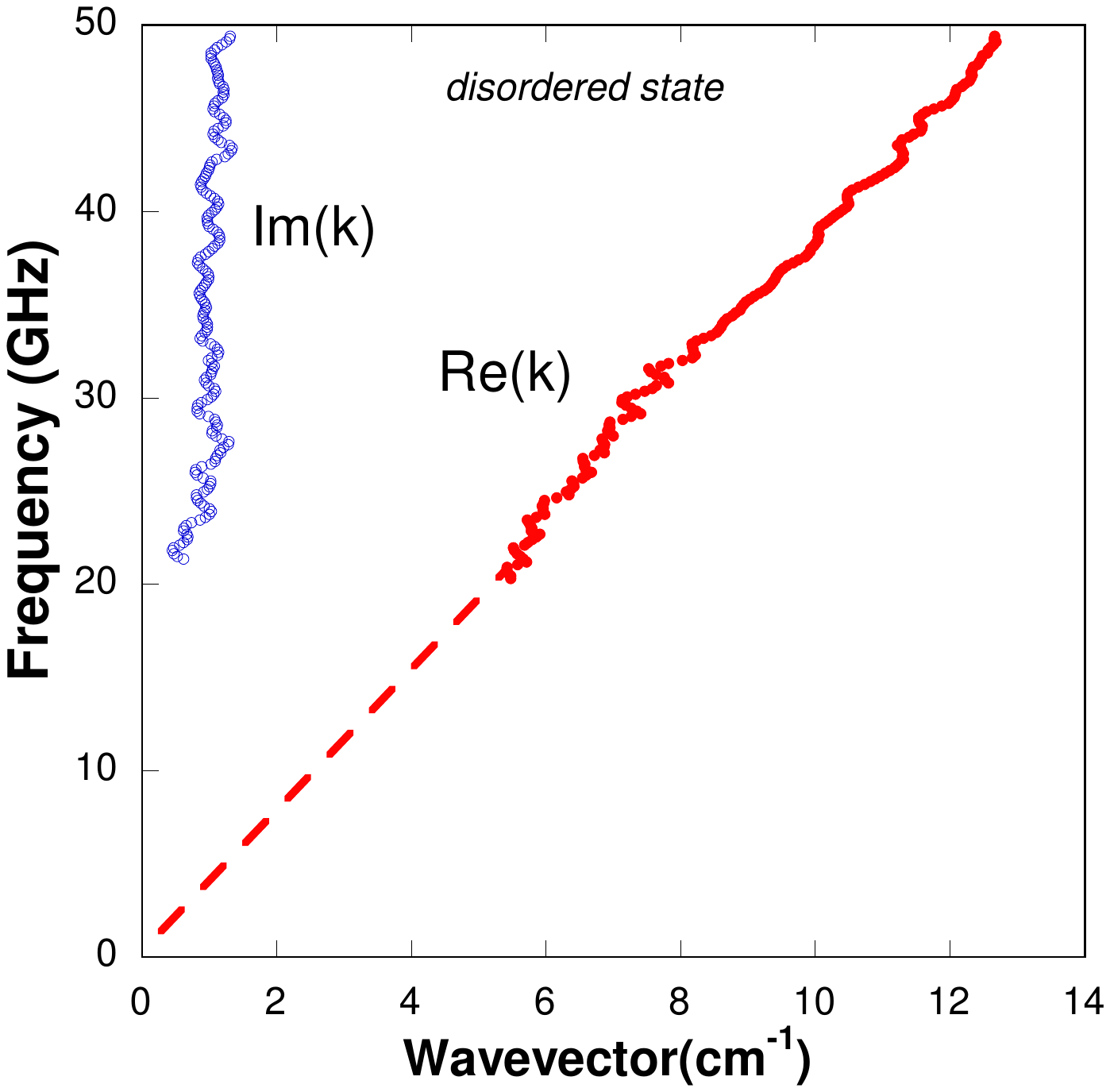}
\caption{Dispersion relations for the ten-layer stack (331 sphere in each layer, interlayer distance is 4.3 mm).   Filled symbols  show $\mbox Re  (k)$  and open symbols show $\mbox Im (k)$, where $k$ is the wavevector. The dashed line shows $k_{0}$- the wavevector in the uniform medium with the same refraction index as our photonic crystal. (a) The ordered state, $H=$160 Oe.  (b) The disordered state, $H=$0. Note the gap between 24 GHz and 36 GHz in the ordered state and its absence in the disordered state. }
\label{fig:7}
\end{figure}

\begin{figure}[ht]
\includegraphics[width=0.8\textwidth]{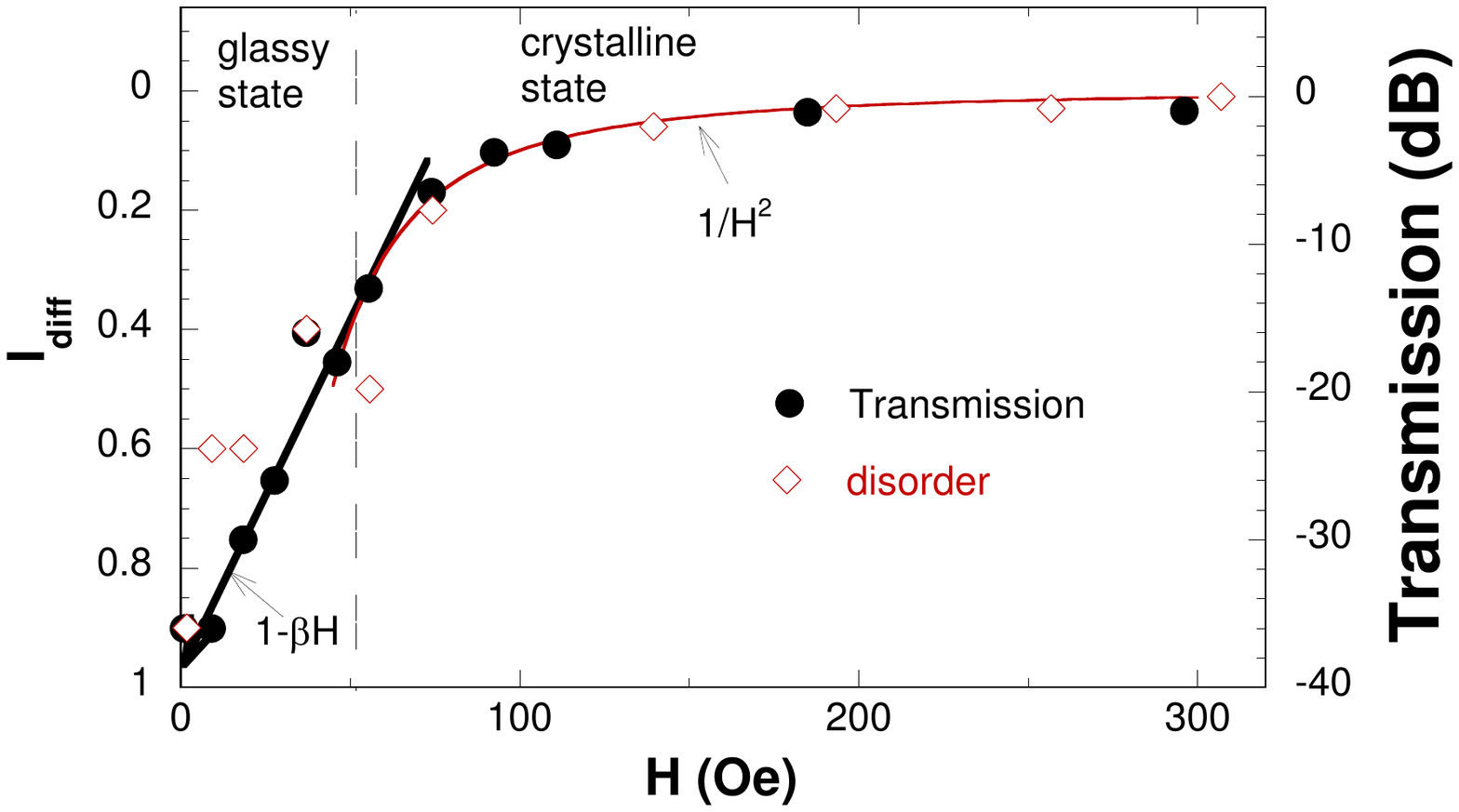}
\caption{ Mm-wave transmission at  46.3 GHz for the sample of  Fig.5 (filled circles) and the intensity of diffuse background (open rhombs), $I_{diff}$, estimated from the Fourier-transform images of Fig.1.  Straight solid line shows linear dependence dependence as predicted by  Eq.\ref{fluct-low} for the glassy state; the thin curved line shows $1/H^2$ dependence as predicted by Eq.\ref{fluctLO} for the polycrystalline state. The  vertical line delineates the  strong field region (where Fourier transform imaqe of Fig.1 shows sharp Bragg peaks, indicating on crystalline state), from the weak field region (here Fourier transform imaqe of Fig.\ref{fig:1} shows sharp diffuse rings  indicating on amorphous or glassy  state.)
}
\label{fig:8}
\end{figure}

\begin{figure}[ht]
\includegraphics[width=0.8\textwidth]{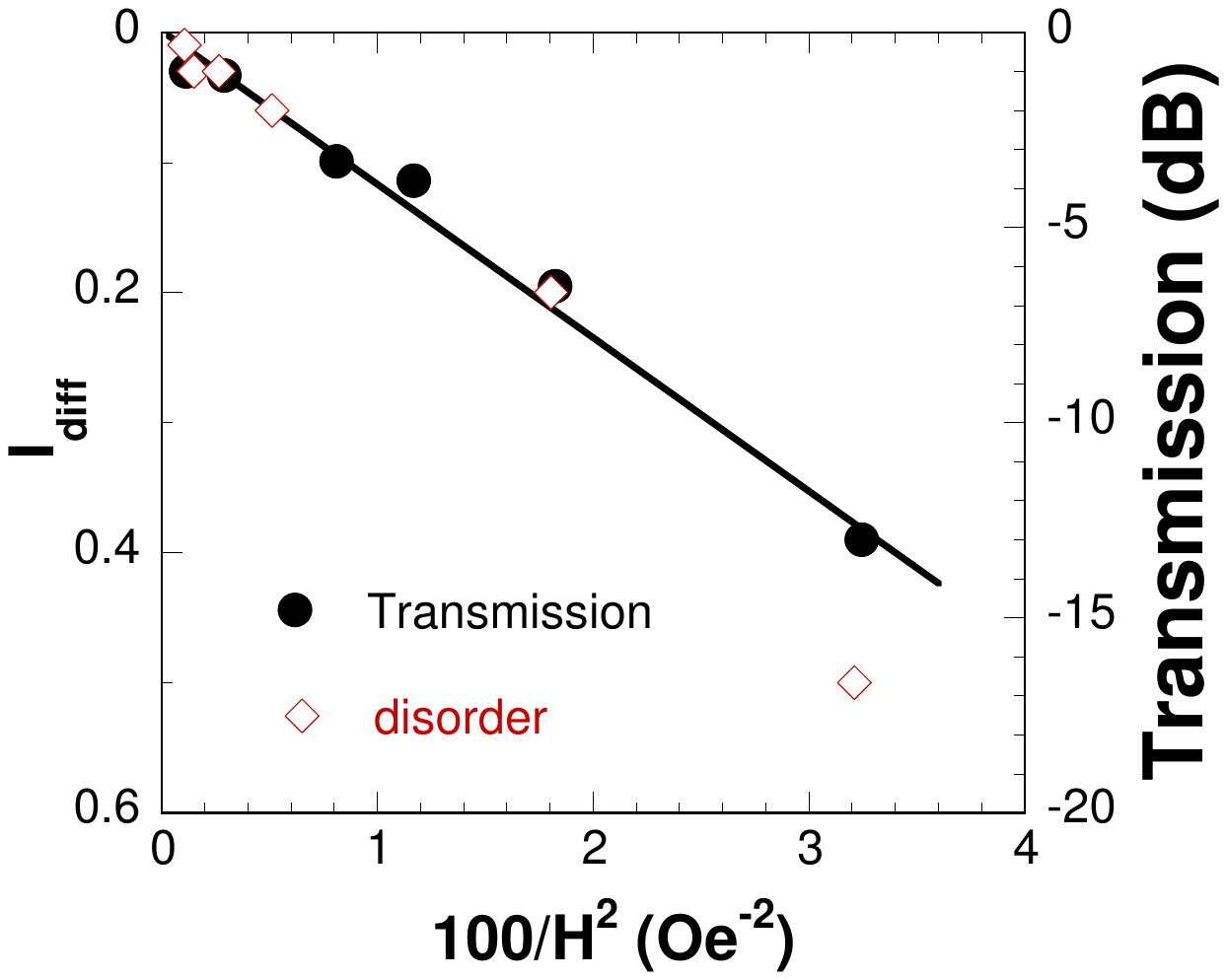}
\caption{ A different representation of the  high-field data of Fig.\ref{fig:8}. Filled circles show mm-wave transmission at  46.3 GHz for the sample of  Fig.5, open rhombs show the magnitude of area density fluctuations $I_{diff}$.  Straight solid line shows linear dependence  on $1/H^2$ as predicted by Eq.\ref{fluctLO}.}
\label{fig:9}
\end{figure}

Since our model accounts fairly well for the magnetic field dependence of the mm-wave transmission, in what follows we only estimate the prefactor $B(f)$ in Eq.\ref{master}. To this end we compare the ratio of transmittances in the completely ordered and in the completely disordered states and for ceratin frequencies. In particular,  for  \textit{antigap},  we substitute into Eq.\ref{antigap}  $N=6$, $f=$46 GHz, $\overline{\delta\rho^{2}}/\rho^2=0.2$,  $a=$ 1.37 mm, $x=0.73$,  $A=0.5$, $n=1.4$, $\rho d n/d \rho=0.7$ and find $\ln T_{disorder}/T_{order}$=-24 dB. This should be compared to the experimentally observed  value of -10-15 dB (Fig.\ref{fig:4}). For the \textit{midgap}, we substitute into Eq. \ref{midgap}, $N=6$, $f=$ 27 GHz, $\overline{\delta\rho^{2}}/\rho^2=0.12$, $x=0.73$, $A=0.5$, $Y=1.9$, $\rho^{2}d^{2} Y/d\rho^{2}=0.4$ and find $\ln T_{disorder}/T_{order}$=-2.5 dB. This should be compared to the observed value of $\sim$-5 dB (Figs.\ref{fig:3},\ref{fig:4}).

 We conclude that our model qualitatively predicts field dependence of the transmission through disordered photonic crystal.  To achieve better quantitative  agreement, there is need  in a more advanced model of the kind suggested by Ref.\cite{Mulholland}. This model should go further than Eqs.\ref{n},\ref{Y},\ref{n1},\ref{Y1} which assume low concentration of scatterers and  Rayleigh scattering regime.  The model should also account for the quadrupole interactions between magnetized particles.

\subsection{Mm-wave transmission through the crystal with a planar defect}
The microwave transmission through the crystal with a planar defect can be treated following the same lines. We consider the photonic crystal split on two halves,  as two photonic crystals  coupled through the Fabry-Perot resonator. Transmission through this device is \cite{Golosovsky}:
\begin{equation}
T_{FP}=\frac{e^{i\frac{2\pi f\Delta}{c}}S_{12}^{2}}{e^{i\frac{4\pi f\Delta}{c}}-S_{11}^{2}}\label{Fabry}
\end{equation}
where $\Delta$ is the spacing between the mirrors and $S_{12}$ and $S_{11}$ are complex transmission and the reflection  coefficients for each of the halves (note that $S_{11}$ and $S_{12}$ are interrelated and both are affected by disorder). The resonant transmission through such a crystal is possible provided the denominator of Eq.\ref{Fabry} is small, i.e. $1-|S_{11}|^2\ll 1$. Even small disorder strongly affects  this condition, hence the transmission through the defect drops dramatically in the disordered state (Fig.\ref{fig:61}).

\section{Discussion and Conclusions}

Continuous variation of electromagnetic properties of our samples  when external magnetic field  goes to zero closely resembles the change of optical properties of liquids upon approaching the critical point.   Indeed, the  optical transmission at the critical point is very small due to enhanced density fluctuations. According to Einstein's theory, the density fluctuations  inversely depend on compressibility which goes to zero at the critical point \cite{LL1}.  The compressibility of our array of magnetizable spheres, $K^{-1}\sim 1/H^2$ (Eq.\ref{mu}), also diverges when $H\rightarrow 0$.  The density fluctuations, as it is shown by Eq.\ref{fluctLO},  increase accordingly. 

Our experiments bear some resemblance to the  studies of electromagnetic wave propagation through ferrofluids \cite{Vlad,ferrofluids1}.  Indeed, in the absence of magnetic field the particles  in ferrofluid are in the disordered state, while in the presence of magnetic field they  self-assemble into chains. This results in field-induced anisotropy which strongly affects the polarization of the electromagnetic wave propagating through such a media. However,  magnetic field does not induce the long-range order in ferrofluids, hence the field-induced anisotropy is \textit{frequency-independent}.  This is very different from our present experiment  where  magnetic field drives the particle array into crystalline state with a long-range order, and this  has a strong \textit{frequency-selective} effect on the microwave transmission. However, two-dimensional ferrofluid layers \cite{ferrofluids2,Hong} bear strong resemblance to our system.

In summary, we demonstrate a metallo-dielectric photonic crystal exhibiting stopband in the mm-wave range.  Mm-wave transmission through this crystal can be controlled by external magnetic field through  magnetic field-induced order-disorder transition. We develop a physical model which describes our experimental data fairly well.   Our concepts may be useful in interpretation of the effect of disorder in photonic bandgap materials. Our results  can be useful for the fabrication of tunable planar photonic crystals, based on surface waves, in particular  surface plasmons. If the surface is covered with the liquid layer containing movable magnetic particles, (especially when this layer represents a photonic crystal with a tunable lattice constant) the propagation of the surface waves can be effectively monitored.  
\begin{acknowledgments}
This work was supported by the VW foundation, Israeli Science foundation, and  Israeli Ministry of Science and Technology. We are grateful to  I. Felner for  magnetization measurements; to A. Frenkel for the help with computer simulations; to B. Laikhtman, L. Schvartsman,  A. Sarychev and V.Freilikher for fruitful discussions. 
\end{acknowledgments}

%%\pagebreak %twocolumn

 \pagebreak%preprint
 
\begin{thebibliography}{}
\bibitem{Joannopoulos}J. D. Joannopoulos, R. D. Meade, J. N. Winn, "Photonic Crystals Molding the Flow of Light ",  Princeton University Press, New Jersey, (1995).
\bibitem{John} K. Busch, S. John, Phys. Rev. E \textbf{58}, 3896 (1998).
\bibitem{Halevi} P. Halevi, F. Ramos-Mendieta, Phys. Rev. Lett. \textbf{85},
1875 (2000).
\bibitem {Kim}S. Kim, V. Gopalan, Appl. Phys. Lett. \textbf{78}, 3015 (2001).
\bibitem{Lyubchanskii} I.L. Lyubchanskii, N.N. Dadoenkova, M.I. Lyubchanskii, E.A. Shapovalov, Th. Rasing, J. Phys. D:Appl. Phys. \textbf{36}, R277 (2003).

\bibitem{Belotelov} V.I. Belotelov, A.K. Zvezdin, J. Opt. Soc. Am. B \textbf{22}, 286 (2002). 

\bibitem{Gates} B. Gates, Y.N. Xia, Advanced Materials \textbf{13},
1605 (2001).
\bibitem{Figotin} A. Figotin, Y.A. Godin, I. Vitebsky, Phys. Rev. B
\textbf{57}, 2841 (1998).
\bibitem{Xu} X.L. Xu, G. Friedman, K.D. Humfeld, S.A. Majetich, S.A.
Asher, Adv. Mater. \textbf{13}, 1681 (2001).

\bibitem{Xu1} C. Xu, X. Hu, Y. Li,X. Liu, R. Fu, J. Zi, Phys. Rev. B \textbf{68}, 193201 (2003).
\bibitem{Bizdoaca} E.L.  Bizdoaca, M.  Spasova, M.  Farle, M.
Hilgendorf, F.  Caruso, Journal of Magnetism and Magnetic Materials
\textbf{240}, 44 (2002).
\bibitem{Saado} Y. Saado, M. Golosovsky, D. Davidov, A. Frenkel,  Phys. Rev. B \textbf{66}, 195108 (2002).

\bibitem{Golos} M. Golosovsky, Y. Saado, D. Davidov, Appl. Phys. Lett. \textbf{75}, 4168 (1999).

\bibitem{Hayes} M.A. Hayes, N.A. Polson, and A.A Garcia, Langmuir \textbf{17}, 2866 (2001).

\bibitem{Ping} Ping Sheng, \textit{Scattering and Localization of classical waves in random media}, World Scientific, Singapore, 1990.

\bibitem {Fan} S.Fan, P.R. Villeneuve, J.D. Joannopoulos, Phys. Rev. B
\textbf{54}, 11245 (1996).
\bibitem{Sigalas1} M.M. Sigalas, C.T. Chan, K.M. Ho, C.M. Soukoulis, Phys.
Rev. B\textbf{ 52}, 11744 (1995).

\bibitem{Sigalas2} M.M. Sigalas, C.M. Soukoulis, C.T. Chan, R. Biswas, K.M. Ho,  Phys. Rev. B \textbf{59}, 12767 (1999).

\bibitem{Freilikher} V.D. Freilikher, B.A. Liansky, I.V. Yurkevich, A.A. Maradudin, A.R. McGurn, Phys. Rev. E \textbf{51}, 6301 (1995).

\bibitem{Deych} L.I. Deych, D. Zaslavsky, A.A. Lisyansky, Phys. Rev. Lett. \textbf{81}, 5390 (1998).

\bibitem{Bayindir} M. Bayindir, E. Cubukcu, I. Bulu, T. Tut, E. Ozbay, C.M.
Soukoulis,  Phys. Rev. B. \textbf{64}, 195113 (2001).

 \bibitem{Asatryan} A.A. Asatryan, P.A. Robinson, L.C. Botten, R.C. McPhedran, N.A. Nicorovici, C. Martijn de Sterke, Phys. Rev. E \textbf{60}, 6118 (1999); \textbf{62}, 5711 (2000).
 
\bibitem{Yannopapas} V. Yannopapas, N. Stefanou, A. Modinos, Phys. Rev. Lett. \textbf{86}, 4811 (2001).

\bibitem{Kaliteevskii} M.A. Kaliteevski, J. Manzanares Martinez, D. Cassagne, J.P. Albert, Phys. Rev. B \textbf{66}, 113101 (2002); Phys. Stat. Sol. (a) \textbf{195}, 612 (2003).

 \bibitem{Zhang} Z. Daozhong, H. Wei, Z. Youlong, L. Zhaolin, C. Bingying, Y. Guozhen, Phys. Rev. \textbf{B 50}, 9810 (1994).
 
\bibitem{KP} A. Kondilis, P. Tzanetakis, Phys. Rev. B \textbf{46}, 15426 (1992); J. Opt. Soc. Am. A \textbf{11}, 1661 (1994).

\bibitem{Sterke} C. Martijn de Sterke, R.C. McPhedran, Phys. Rev. B 
\textbf{47}, 7780 (1993).

\bibitem{Bulgakov} S.A. Bulgakov and M. Nieto-Vesperinas, Waves in 
random media, \textbf{7}, 183 (1997); J. Opt.Soc.Am. A \textbf{15}, 503 (1998).

\bibitem{Kondilis} A. Kondilis, Phys. Rev. B \textbf{55}, 14214 (1997).

\bibitem{Genack}  A.A.Chabanov, M.Stoytchev, A.Z.Genack, Nature \textbf{404},
850 (2000); M. Stoytchev, A.Z. Genack, Phys. Rev. B \textbf{55}, R8617 (1997).

\bibitem{Netti} M.E. Zoorob, M.D.B. Charlton, G.J. Parker, J.J. Baumberg, M.C. Netti, Nature \textbf{404}, 740 (2000).

\bibitem{Li} Z.Y.Li and Z.Q. Zhang, Adv. Mat. \textbf{13}, 433 (2001).

\bibitem{Levy} O. Levy, Phys. Rev. E \textbf{61}, 5385 (2000).

\bibitem{Golosovsky} M. Golosovsky, Y. Neve-Oz, D. Davidov, A. Frenkel, Phys. Rev. B \textbf{70}, 115105 (2004).

\bibitem{Ghazali} A. Ghazali, J.C. Levy, Phys. Rev. B \textbf{67}, 064409 (2003). 

\bibitem{Kusner} R.E. Kusner, J.A. Mann, A.J. Dahm, Phys. Rev. B \textbf{51}, 5746 (1995).

\bibitem{Golos-synt} M. Golosovsky, Y. Neve-Oz, D. Davidov, Synth.Metals \textbf{139}, 705 (2003).


\bibitem{LO} A.I. Larkin and Yu. N. Ovchinnikov, J. of Low-Temp. Phys., \textbf{34}, 409 (1979).

\bibitem{LL} L. D. Landau and E.M. Lifshitz, \textit{Statistical Physics}. Pergamon, New York, 1980.

\bibitem{Portis} A.M. Portis, \textit{Electromagnetic fields: Sources and media}, J. Wiley, N.Y., 1978, ch.14. p.562.

\bibitem{Tretyakov} S.A. Tretyakov, A.J. Viitanen, S.I. Maslovski, I.E. Sarela, IEEE Trans. on Antenna and Propagation, \textbf{51}, 2073 (2003).

\bibitem{Hulst} H.C. van de Hulst, \textit{Light scattering by small particles}, J. Wiley, N.Y. 1937.

\bibitem{Bohren} C.F. Bohren and D.R. Huffman, \textit{Absorption and scattering of light by small particles}, Wiley, N.Y. 1998.

%\bibitem{bohren1} C.F. Bohren, Journal of the Atmospheric science, \textbf{43}, 468 (1986).
\bibitem{A} The local field factor in dipole arrays is closely related to the dipole energy per site. In particular, for highly symmetrical three-dimensional  dipole arrays, $A=1$. For the planar  dipole array, the local field factor depends on the orientation of the dipoles with respect to the array plane. For the out-of-plane  orientation, the lattice sum calculations \cite{Kusner,Ghazali} yield   $A= -1.037$ for the hexagonal lattice, and $A= -1.053$ for the square lattice. For the in-plane  orientation, $A_{sq}= 0.56$  for the square lattice \cite{Tretyakov}. We are not familiar with the corresponding calculation for the hexagonal dipole array, but the estimate accounting for  the contribution from the close neighbors suggests $A_{hex}\approx 0.5$. Since the local field factor is different for the square and hexagonal lattice, it should be sensitive to shear deformations.


\bibitem{Pendry} J.B. Pendry, Adv. Phys. \textbf{43}, 461 (1994). 


\bibitem {Yeh}P. Yeh, \textit{Optical waves in layered media}, J.Wiley, N.Y. 1988.

\bibitem{LL1} L. D. Landau and E.M. Lifshitz, \textit{Electrodynamics of continuous media}, Pergamon, New York, 1975. 


\bibitem{Goldstein} G.D. Cody, T. Tiedje, B. Abeles, B. Brooks, Y. Goldstein, Phys. Rev. Lett. \textbf{47}, 1480 (1981).

\bibitem{Romanov} S.G. Romanov, C.M. Sotomayor Torres, Phys. Rev. E \textbf{69}, 046611 (2004).


\bibitem{Lozovik} V.M. Bedanov, G.V. Gadyak, Yu.E. Lozovik, Phys. Lett. \textbf{92A} ,400 (1982).

\bibitem{Zahn} K.Zahn, R. Lenke, G. Maret, Phys. Rev. Lett. \textbf{82}, 2721 (1999).
\bibitem{Bubeck}R. Bubeck, C. Bechinger, S.Neser, P. Leiderer, Phys. Rev. Lett. \textbf{82}, 3364 (1999).

\bibitem{Weiss}J.J. Weiss, J. Phys.: Condens. Matter \textbf{15}, S1471 (2003).

\bibitem{Peeters} M. Kong, B. Partoens, F.M. Peeters, Phys. Rev. E \textbf{67}, 021608 (2003).


\bibitem{Mulholland} G.W. Mulholland, C.F. Bohren, K.A. Fuller, Langmuir, \textbf{10}, 2533, (1994).

\bibitem{Vlad}T.H. Ji, V.G. Lirtsman, Y. Avny, D. Davidov,
Advanced Materials \textbf{13}, 1253 (2001).
\bibitem{ferrofluids1} J.E.Martin, K.M. Hill, C.P. Tigges, Phys.
Rev. E, \textbf{59} 5676 (1999).
\bibitem{ferrofluids2} S.Y.Yang, Y.P.Chiu, B.Y. Jeang, H.E. Horng,
C.Y. Hong, H.C. Yang, Appl. Phys. Lett., \textbf{79}, 2372 (2001).

\bibitem{Hong} C.Y. Hong, Y.S. Yeh, S.Y. Yang, H.E. Horng, H.C. Yang, J.  Magn. Magn. Mat. \textbf{283}, 22 92004).

\bibitem{Synt} In our preliminary study  \cite{Golos-synt} we made an attempt to explain  our results  by Eq.\ref{fluct-uni}, assuming  a single crystalline particle array. In fact, our results are better described by Eq.\ref{fluctLO} which assumes a polycrystalline particle array.


 \end{thebibliography}
\end{document}